\documentclass{article}
\usepackage{ijcai24}

\usepackage{times}
\usepackage{soul}
\usepackage{url}
\usepackage[hidelinks]{hyperref}
\usepackage[utf8]{inputenc}
\usepackage[small]{caption}
\usepackage{graphicx}
\usepackage{amsmath}
\usepackage{amsthm}
\usepackage{booktabs}
\usepackage{algorithm}
\usepackage{algorithmic}
\usepackage[switch]{lineno}
\usepackage{tabularx}

\title{Towards Proactive Interactions for In-Vehicle Conversational Assistants Utilizing Large Language Models}

\author{
Huifang Du$^1$
\and
Xuejing Feng$^1$ \and
Jun Ma$^{1}$\and
Meng Wang$^1$\and
Shiyu Tao$^2$\and
Yijie Zhong$^1$\and
Yuan-Fang Li$^3$\And
Haofen Wang$^1$\\
\affiliations
$^1$Tongji University\\
$^2$Beijing Technology and Business University\\
$^3$Monash University\\
\emails
\{duhuifang, fengxuejing, jun\_ma, mengwangtj\}@tongji.edu.cn,\\
\{dun.haski, carter.whfcarter\}@gmail.com, \\
ms\_taoshiyu@163.com,\\
yuanfang.li@monash.edu
}

\begin{document}

\maketitle

\begin{abstract}
Research demonstrates that the proactivity of in-vehicle conversational assistants (IVCAs) can help to reduce distractions and enhance driving safety, better meeting users' cognitive needs. However, existing IVCAs struggle with user intent recognition and context awareness, which leads to suboptimal proactive interactions. Large language models (LLMs) have shown potential for generalizing to various tasks with prompts, but their application in IVCAs and exploration of proactive interaction remain under-explored. These raise questions about how LLMs improve proactive interactions for IVCAs and influence user perception. To investigate these questions systematically, we establish a framework with five proactivity levels across two dimensions—assumption and autonomy—for IVCAs. According to the framework, we propose a ``Rewrite + ReAct + Reflect'' strategy, aiming to empower LLMs to fulfill the specific demands of each proactivity level when interacting with users. Both feasibility and subjective experiments are conducted. The LLM outperforms the state-of-the-art model in success rate and achieves satisfactory results for each proactivity level. Subjective experiments with 40 participants validate the effectiveness of our framework and show the proactive level with strong assumptions and user confirmation is most appropriate.

\end{abstract}

\section{Introduction}
In-vehicle conversational assistants (IVCAs) are an integral component in smart cockpits and play a vital role in facilitating human-agent interaction \cite{lee2022systematic}. They can deliver features including navigation, entertainment control, and hands-free phone operation \cite{braun2019your}. Despite the promising prospects and strides made in IVCAs' development, the proactive interactions from a human-centered perspective in the vehicle context are relatively limited. For example, existing IVCAs mostly passively receive and execute simple commands \cite{meck2023may,meck2021design,lin2018adasa}, though the proactive interaction concepts are proposed in some car manufacturers \footnote{https://www.mercedes-benz.de/passengercars/technology/mbux-zero-layer.html}. The issues above can be approached from two angles. \textbf{Firstly, current research lacks a clear and helpful definition of proactivity for IVCAs. Secondly, there are technical limitations to achieving satisfactory proactive interactions, such as poor intent recognition \cite{mi2022cins} and context awareness \cite{shen2022kwickchat}}. As the demand for better interaction experiences rises, IVCAs are expected to manage complex tasks and offer proactive support \cite{volkel2021eliciting}. Particularly, through proactively providing information and addressing the anticipated issues\cite{kim2020interruptibility}, IVCAs can effectively offer personalized services and reduce driver cognitive load, thus improving driving safety and experience.

\begin{figure}
        \centering
        \includegraphics[height=0.25\textwidth]{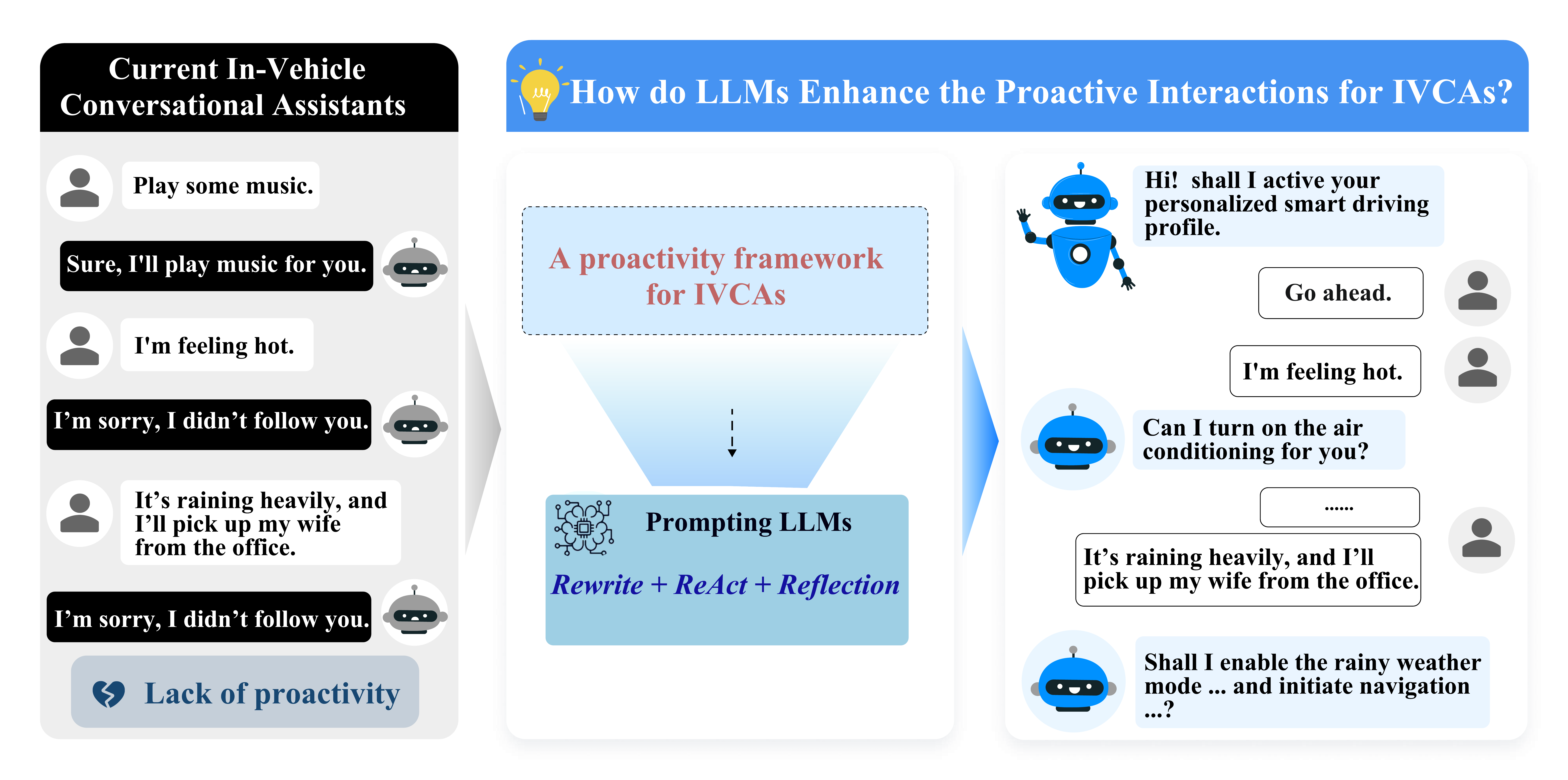}
        \caption{The motivation and main contributions of our work to exploring effective proactive interactions for IVCAs based on LLMs.}
        \label{fig: teaser}
    \end{figure}

From the perspective of human-computer interaction (HCI) research, proactive behaviors of conversational assistants can be summarized into two essential elements \cite{peng2019design,grant2008dynamics,parker2010taking}: \emph{autonomy} (the capability to execute intended tasks), and \emph{assumption} (anticipation of users needs). Many efforts center around these two dimensions. For instance, the Interface-Proactivity (IP) continuum was proposed to define five different proactivity levels of autonomy, ranging from zero to full autonomy \cite{isbell2005ip}. Building upon the IP continuum, proactive dialogues are categorized into four levels: None, Notification, Suggestion, Intervention \cite{kraus2021role,kraus2020effects}. Besides, a three-level proactivity policy framework for decision-making support assistants was defined across the assumption and autonomy dimensions~\cite{peng2019design}. Yet, how IVCAs in driving contexts proactively interact with users is still an open issue. Some studies try to understand the impact of proactivity on human-vehicle interaction from the viewpoint of interruptions \cite{kim2020interruptibility,cha2020hello}. They provide valuable case studies, but mainly focus on the timing and linguistic impacts, without offering comprehensive interaction strategies.

As for conversational technologies, extensive research has been conducted in the general domain \cite{young2022fusing}, inspiring the studies of IVCAs. 
Still, conversation support for IVCAs, such as Adasa \cite{lin2018adasa} and CarExpert \cite{rony2023carexpert}, have mainly focused on providing driving-related knowledge with data derived from user manuals. In contrast, we try to address users' task-related requests in various driving scenarios. Furthermore, current dialogue models typically require large amounts of labeled data, incur high costs, and may not generalize well to other tasks. Recently, large language models (LLMs), such as GPT-4 \footnote{https://arxiv.org/abs/2303.08774} and Vicuna \footnote{https://lmsys.org/blog/2023-03-30-vicuna/}, have shown impressive understanding and generation capabilities to many tasks with prompts (i.e.\ without updating model parameters). The ability to learn from limited data is highly advantageous, but few studies have been conducted to understand their viability for IVCAs. 

In this paper, we investigate enhancing proactive interactions between users and IVCAs by mitigating the aforementioned two issues: the lack of clear proactivity definition for IVCAs and technical limitations. To thoroughly explore proactive interactions of IVCAs, it is imperative to establish a formal formulation and ensure a consistent implementation between IVCAs and users. Drawing on previous research \cite{peng2019design,isbell2005ip,kraus2021role,kraus2020effects}, we build a proactivity framework with five levels across assumption and autonomy dimensions while incorporating user control as a design constraint. Based on the framework, we investigate LLMs' feasibility in achieving different levels of proactivity for IVCAs. We prompt LLMs by proposing a ``Rewrite + ReAct + Reflect'' approach to get a response. Specifically, we first rewrite casual questions of users to be more normal in driving contexts. Then we not only prompt LLMs to reason by the proactive interaction instructions and search for external supportive knowledge but also make them reflect on whether the generated answers fulfill the designated level of proactivity. Our work also provides insights into how LLMs can integrate various in-vehicle information for understanding and decision-making tasks. 

We extensively experiment with the LLM model gpt-3.5-turbo to investigate its capability to achieve various levels of proactivity for IVCAs. Results show that the LLM not only achieves a superior success rate (93.72\%) than the state-of-art models for task-oriented dialogue but also satisfactory proactivity attainment rates for each proactivity level (more than 78\%). Furthermore, we explore the effects of different levels of proactive interactions on human perception with 40 participants. For a more realistic setting, we develop an IVCA simulator based on the LLM to implement an actual conversation environment. Experimental results indicate that the proactivity level with strong assumptions and user confirmation is most preferred. As it offers natural and helpful assistance and user confirmations, it's considered the most appropriate. Notably, our work is the first to explore proactive interactions for IVCAs using LLMs, verifying the potential of LLMs for IVCAs. In summary, the main contributions are as follows: 

\begin{itemize}
  \item We establish a proactivity framework for IVCAs with five levels along the dimensions of assumption and autonomy while integrating user control as a design principle. The framework lays a theoretical foundation for systematically studying proactive interactions for IVCAs.
  
  \item To our knowledge, we are the first to investigate the potential of LLMs in improving proactive interaction experiences for IVCAs. We utilize a ``ReAct + Reflect'' strategy to prompt LLMs to achieve various levels of proactivity with satisfactory performance.
  
  \item Comprehensive experimental results show that our approach is feasible to enhance the interaction experience for users. We observe that proactivity significantly influences user perceptions and users prefer proactive interactions with strong assumptions and user control.
\end{itemize}

\section{RELATED WORK}
The evolution of IVCAs has been a significant research subject within the context of human-computer interaction (HCI) and artificial intelligence (AI). This section explores related work in the areas of proactive interaction, prompting LLMs.

\subsection{Proactivity of the Intelligent Assistants}
Proactivity is determined by the following two factors: assumption, and autonomy in the domain of occupational and organizational psychology \cite{grant2008dynamics,parker2010taking}. Based on the two elements, proactive behaviors of assistants are often discussed in HCI \cite{peng2019design,kraus2021role}. Among these studies, the challenges of \textit{if}, \textit{how}, and \textit{when} to take proactive action for dialogue assistants are proposed \cite{nothdurft2014finding} and become the guidelines for designing proactive assistants then. The \textit{if} question stresses the necessity. Many studies demonstrate that proactive behaviors of an assistant system affect the user's perception \cite{peng2019design,kraus2020effects} and proactivity is considered one of the users' desired features for perfect assistants \cite{zargham2022understanding}. Regarding the \textit{how} research, some works give examples to answer the \textit{how} question \cite{zargham2022understanding,meck2023may}, but they mainly focus on specific features like linguistic styles, tone of voice, gestures, etc. There are also some works developing guidelines for general dialogues between humans and assistants \cite{isbell2005ip,peng2019design,kraus2020effects,kraus2021role}. As for the research question of \textit{when}, some studies try to find the balance between being helpful and being intrusive decided by proactivity from the viewpoint of interruptions \cite{kim2020interruptibility} and linguistic impacts \cite{cha2020hello}. In this paper, we focus on building proactive interaction strategies tailored for IVCAs to respond to the \textit{how} research question, which also lays the groundwork for opportune interactions.

\subsection{Prompting Large Language Models}
Recently, large language models (LLMs) \cite{brown2020language} have shown emergent abilities \cite{schaeffer2023emergent} and have led to a new paradigm in creating natural language processing systems. Unlike traditional methods that rely on a well-selected, labeled training dataset, LLMs have introduced a new technique, prompt engineering. In-context learning (ICL), prompting LLMs with a few examples \cite{dong2022survey}, can generalize to various tasks like summarization, question answering, and code generation without updating parameters. ICL is adopted in our study to transform users' diverse and casual expressions into formal questions.
More Helpful prompting techniques are proposed to interface with LLMs \cite{wei2022chain,yao2023tree,yao2022react}. For example, chain of thought (CoT) \cite{wei2022chain} shows intermediate reasoning steps of the examples to boost the prompting performance. Using the technique of Tree of Thoughts (ToT), LLMs can make thoughtful decisions by considering many different reasoning paths and self-evaluating options \cite{yao2023tree}. The ReAct prompting framework leverages LLMs to produce reasoning traces and task-specific actions while enabling the collection of external information \cite{yao2022react}. It also enhances the trustworthiness and interoperability of LLMs by using the problem-solving process. We adopt ReAct prompting in our work to incorporate external knowledge and implement proactive interaction strategies. Additionally, we include a reflective function at the end to ensure that LLMs align with the desired level of proactivity.


\section{Design of Proactive Interaction Strategies} \label{Strategies}
In this section, we formulate proactive interaction behaviors reflecting the unique characteristics of interacting with IVCAs in driving contexts. Specifically, we apply the concept of proactivity, originally from the field of occupational and organizational psychology \cite{grant2008dynamics,parker2010taking}, to the domain of HCI \cite{peng2019design}, considering two essential factors: autonomy and assumption. The first factor, system autonomy, which refers to the ability to perform tasks without direct user commands, has been the subject of study in various earlier works. These include the autonomy scale definition in \cite{rau2013effects}, the IP continuum in \cite{isbell2005ip}, the three-level proactivity framework in \cite{peng2019design}, and four-level proactivity in \cite{kraus2020effects}. We follow the principles of autonomy as outlined in these works when designing the proactive behaviors of IVCAs. Regarding the system assumption, it is closely associated with the ability to anticipate users' potential intentions \cite{kraus2020effects}. Many methods utilize human actions or poses, such as gaze and body positioning, to make predictive inferences. We attempt to make assumptions by comprehending the driver's utterances in driving contexts. Building upon prior work, we design the proactivity scales for IVCAs based on assumptions and autonomy as well. However, considering the direct implications of IVCAs on driving safety and user experience, we particularly account for the importance of user control \cite{kraus2021role,kraus2020effects}. We incorporate user control as a design principle or constraint, dividing the levels of proactivity into five levels based on assumptions and autonomy. Within each proactive level, we discuss the degree and manner of user control. The proactive interaction guidelines at the five levels are derived as follows:


    \textbf{Level 1}. At this level, IVCAs make no assumptions and passively receive and execute the user's instructions. The user has full control over the behavior of IVCAs, and IVCAs will not take any action without instructions. For instance, ``Driver: Please turn on the air conditioner. IVCAs: Sure.''

    \textbf{Level 2}. IVCAs at this level demonstrate some assumptions, which means IVCAs make preliminary judgments based on limited utterance information. They may identify potential issues or suggest possible solutions based on the assumptions. However, they rely on the user's confirmation before taking any proactive steps. For example, ``Driver: I'm feeling hot. IVCAs: Shall I activate the air conditioning for you? Driver: Go ahead.''


    \textbf{Level 3}. IVCAs at this level show the same level of assumption ability as level 2. However, they can automatically take actions with minimal user inputs during the interaction, and they will execute actions based on these inputs. For instance, ``Driver: I'm feeling hot. IVCAs: I will activate the air conditioning for you. How about 25 degrees Celsius okay? Driver: Sounds good. Thanks''

    \textbf{Level 4}. At this level, IVCAs become highly adaptive, making assumptions based on extensive historical data and deep learning of user behavior. They may initiate conversations and offer suggestions, like providing personalized entertainment options and adjusting responses according to user preferences. However, users retain the right to confirm or adjust proposals before execution. For example, ``IVCAs: Would you like me to set the air conditioning to your preferred temperature of 25 degrees Celsius? Driver: Yes, that would be helpful. IVCAs: The temperature has been set.''

    \textbf{Level 5}. IVCAs are adaptive at this level with strong assumptions like the level at 4. Additionally, they have high autonomy to execute their assumptions automatically with some explanations. However, users can still intervene to stop execution. For example, ``IVCAs: You're in the car. I'll adjust the air conditioning to your preferred temperature of 25 degrees Celsius. Driver: No, thanks.''

Our proactive interaction framework, clearly delineating five levels of proactivity, provides more specific guidance for the design of IVCAs. Additionally, the framework can serve as a benchmark for evaluating the level of proactivity in existing IVCAs, aiding in identifying shortcomings in current systems and guiding future improvement directions. Leveraging this framework, we conduct user studies to discover which level of proactivity in IVCAs is most appropriate.



\section{Rewrite + ReAct + Reflect Prompting} \label{reflection}
\begin{figure*}
        \centering
        \setlength{\belowcaptionskip}{-5 mm}
        \setlength{\abovecaptionskip}{0.2 cm}
        \includegraphics[height=0.32\textwidth]{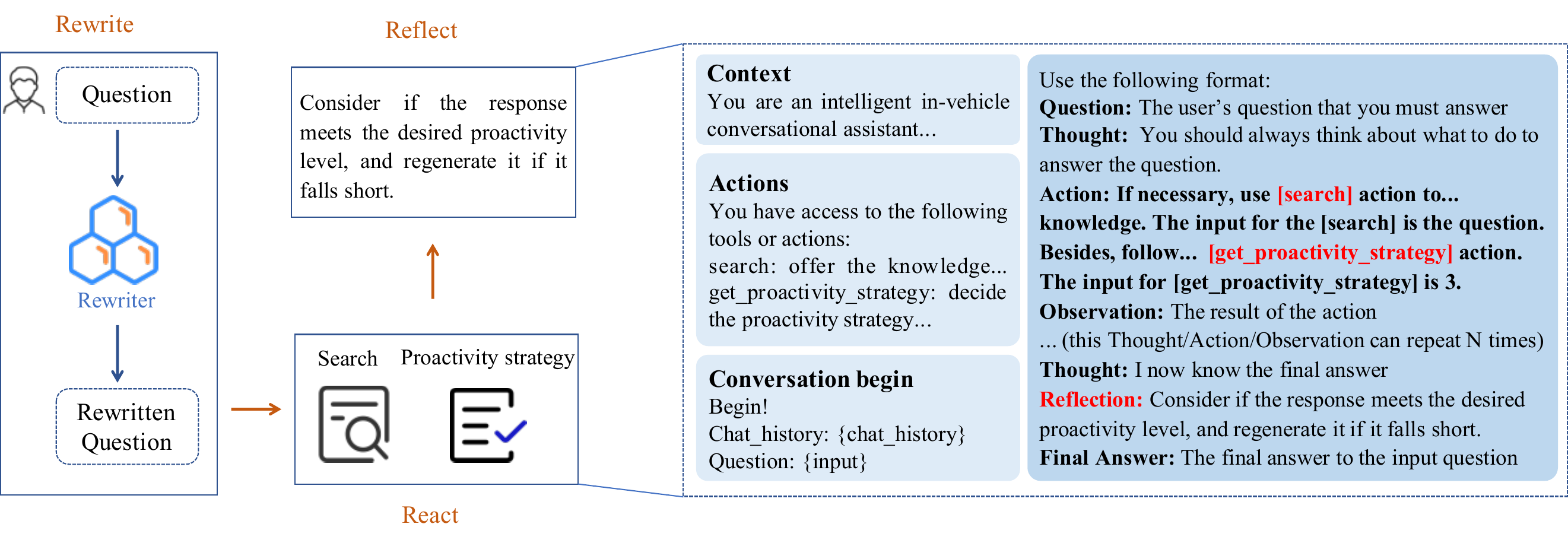}
        \caption{The overview of ``Rewrite + ReAct + Reflect'' prompting. The right is an example prompt based on the ReAct framework incorporating the rewritten question and reflection step.}
        \label{fig: prompt}
    \end{figure*}

In this section, we answer the question of ``How to prompt LLMs to achieve accurate dialogue task completion and align with different levels of proactivity for IVCAs?''. We give an introduction to the task and our prompting strategy. The overview of our ``Rewrite + ReAct + Reflect'' architecture is shown in Figure \ref{fig: prompt}.

\subsection{Task Formulation}
We focus on achieving task-oriented conversations at the formulated proactivity levels in vehicles
by prompting LLMs. Given the dialogue history $H_{t} = (q_{1}, r_{1}, ..., q_{t-1}, r_{t-1})$ and the user's current utterance $q_{t}$, we aim to get the correct response $P_{llm}(y|H_{t},q_{t})$ using question rewrite, ReAct prompt \cite{yao2022react} and the final reflect stage to identify and correct any biases or uncertainties in understanding the proactive interaction strategy in the response. 

\subsection{Question Rewriting}
During conversations with IVCAs, users express themselves in diverse and casual ways, such as saying, ``The smell in the car is a bit pungent.'' To facilitate user-centered interactions, IVCAs should be capable of understanding and responding to various natural language expressions of users in driving contexts. To enhance the accuracy of completing dialogue tasks, it is necessary to align the user's natural inputs with the semantic space of the in-vehicle knowledge bases or contexts. LLMs have powerful language comprehension capabilities and we prompt the LLM with the in-context learning (ICL) technique to convert users' expressions into in-vehicle task-oriented questions. We utilize a few examples to help the LLM gain a deeper understanding of the rewriting question tasks. For instance, ``The smell in the car is a bit pungent'' is transformed into ``Activate the car's fresh air circulation mode.'' An example prompt is shown in the Appendix \ref{QR}.

\subsection{ ReAct + Reflect}
After rewriting the user's question, we improve proactive interactions for IVCAs using the ReAct + Reflect prompting strategy. ReAct \cite{yao2022react} prompts LLMs to trace reasoning and then execute task-specific actions, allowing the model to integrate external knowledge. The term ``actions'' refers to the functions that LLMs can employ. We include \textbf{search}[\textit{question}] and \textbf{get\_proactivity\_strategy}[\textit{number}] as the actions. The search action retrieves the most relevant knowledge with the rewritten question from the knowledge vector store to better support the conversation. The embedding model \footnote{https://github.com/FlagOpen/FlagEmbedding} is used to vectorize the rewritten question and the knowledge in databases. The get\_proactivity\_strategy action prompts the LLM to achieve the desired level of proactivity according to the strategies mentioned above. The search action takes the rewritten question as input, while get\_proactivity\_strategy takes the proactivity level number as input to get a specific proactive interaction strategy as illustrated in Section \ref{Strategies}.

Additionally, some studies suggest that due to limitations in the model's memory capacity, LLMs could forget preceding information as the length of the prompt increases \cite{lu2020understanding}. In our work, the multiple reasoning and retrieved knowledge may lead to an excessively lengthy prompt, hindering LLMs from achieving the precise proactive level. As a solution, we implement a ``reflect'' stage before generating the final response. This stage encourages LLMs to assess whether their response aligns with our chosen proactive strategy and, if not, to regenerate the answer.

\section{Capability Experiments} \label{experiment1}
We conduct experiments to verify the feasibility of using LLMs to improve IVCAs in proactive interactions. 

\subsection{Data Collection} 
We follow the data format of the In-Car dataset \cite{eric-etal-2017-key} to construct multiple knowledge bases covering various scenarios. The In-Car dataset includes weather inquiries, calendar planning, and navigation data. We extend the dataset with in-car functions, environmental conditions, and user profiles. Fields for the knowledge base of each scenario are designed as comprehensively as possible to ensure they address the potential questions users may have within these scenarios. Based on the knowledge bases, we finally obtain a total of 1,302 queries. The covered scenarios and knowledge base examples are shown in Appendix \ref{DC}.



\subsection{Experimental Setups} 
We evaluate whether LLMs, prompted with our designed prompts, can reach each proactivity level with high quality using two metrics: \textit{success rate} (the percentage of successfully achieved user requests or tasks within conversations) and \textit{proactivity attainment rate} (the proportion of LLMs reaching the required level of proactivity). We employ the LLM gpt-3.5-turbo in our experiments. For \textit{Success rate}, we compare the results of gpt-3.5-turbo with the state-of-the-art model TSCP \cite{lei2018sequicity}, LABES \cite{zhang2020probabilistic}, Galaxy\cite{he2022galaxy} for task-oriented dialogue. TSCP is a sequence-to-sequence model with belief spans to track dialogue context. LABES is a dialog model that uses unlabeled data to improve belief state tracking. GALAXY leverage semi-supervised learning to improve the dialogue performance. TSCP, LABES, and Galaxy were all fine-tuned on the training set of the In-Car dataset. As for the proactivity attainment rate, we follow the evaluation method \cite{sun2023safety} to conduct scoring statistics:
$$
Rate =\frac{\sum_{q \in Q} I(C=n)}{N_{Q}} \times 100 \%, \eqno{(1)}
$$where $Rate$ denotes the percentage of the scores labeled as $n$. $n$ is from 1 to 5 in line with the proactivity levels. $C$ means the conversation generated by the LLM. Besides, $Q$ is the collected questions, and $N_{Q}$ is the number of the queries.




\subsection{Result Annotation and Analysis} 
LLMs are generative models and may generate different expressions with correct answers. Therefore, we adhere to the common practice of evaluating language generation quality using human ratings. We assign a 0 when the task is not completed by the IVCAs. As for proactivity, we utilize scoring scales from 1 to 5, representing different proactivity levels according to the principles and rules outlined in our framework. We involve three experts, none of whom are the authors of this paper, to annotate the experimental results. Two specialize in computer science, and the third is from the HCI field. We ask them to rate every dialogue independently. When the annotators assign different scores for the same dialogue, the majority principle is used to resolve the inconsistencies. When all three are different, we discard the dialogue directly, resulting in 1,275 dialogues. The conversation counts of each proactivity level are shown in Table \ref{tab: conversation_number}.

\begin{table}\small
\caption{The conversation statistics for each proactivity level.}
\label{tab: conversation_number}
\centering
\tabcolsep=0.1cm
\renewcommand{\arraystretch}{1.3} 
{
\begin{tabular}{cccccc}
\hline
Conversations & Level 1 & Level 2  & Level 3  & Level 4  & Level 5 \\
    \hline
    number & 210 & 301   &  179  & 177  & 408 \\
    \hline
\end{tabular}
}
\end{table}

After human annotation, we get the experimental results. As shown in Table \ref{tab: success_rate}, gpt-3.5-turbo achieves the success rate (93.72\%), greatly outperforming the other mo. TSCP, LABES, and Galaxy struggle to respond to users' naturally expressed demands, such as ``I'm feeling hot'' while this is easily manageable for LLMs (as shown in Figure \ref{fig: teaser}). This also highlights the capability of LLMs to anticipate user intent within the assumption dimension. The score distribution in Table \ref{tab: proactivity_rate} illustrates that using our ``Rewrite + ReAct + Reflect'' prompts for various proactivity levels, over 78\% of the conversations receive scores within the anticipated ranges. Besides, it can be observed that the results of ``Rewrite + React + Reflect'' are superior to those obtained with the ReAct strategy alone (in parentheses), confirming the effectiveness of the Reflect stage. 

The reasons why some conversations are beyond the expected levels may be because of (1) Context influence. Some studies \cite{MishraKBH22} indicate instructions within a context significantly impact LLMs, while there is no definitive conclusion about the best way to formulate the instruction. (2) Organization of demonstration. Prompt engineering depends on how demonstrations are organized \cite{MishraKBH22}. It's important to manage the demonstration format, the number, and the order of demonstration examples. We need to further explore these points in our future work.


\begin{table}\small
\caption{Performance of different models on success rate.}
\label{tab: success_rate}
\centering
\renewcommand\arraystretch{1.3}
{
\begin{tabular}{lc}
\hline
Model & Success rate \\
    \hline
    TSCP & 64.32 \\
    LABES & 61.60 \\
    GALAXY & 69.00 \\
    gpt-3.5-turbo & \textbf{93.72} \\
    \hline
\end{tabular}
}
\end{table}

\begin{table*}\small
\caption{The proactivity attainment rates at each level. The numbers on the left indicate the specific proactivity strategy used in the prompt, while the percentage on the right represents the proportion reaching each proactivity level. Values in parentheses show outcomes from the ReAct strategy without Reflect stage.}
\label{tab: proactivity_rate}
\centering
\renewcommand{\arraystretch}{1.3} 
{
\begin{tabular}{cccccc}
\hline
Strategies & Level 1 & Level 2  & Level 3  & Level 4  & Level 5 \\
    \hline
    1 & \textbf{87.88 (84.90)} & 0 (3.01)   &  12.12 (12.09)  & 0  & 0 \\
    2 & 6.32 (5.81) & \textbf{78.25 (77.92)}   &  15.44 (16.27)  & 0  & 0 \\
    3 & 1.18 (1.87)  & 16.57 (18.01)  &  \textbf{82.25 (80.12)}  & 0  & 0 \\
    4 & 0 & 0   &   0 (0.08)  & \textbf{90.12 (87.69)}  & 9.88 (12.23) \\
    5 & 0.54 (0.14) & 0.54 (0.62)   &   0 (0.32)  & 17.79 (18.09)  & \textbf{81.13 (80.83)} \\
    \hline
\end{tabular}
}
\vspace{-1 em}
\end{table*}

\section{Subjective Experiments}
Drawing from the capability experiments, it is clear that LLMs exhibit competence in dialogue comprehension and proactive interaction. In this section, we focus on validating our proactivity framework for IVCAs and assessing the effects of the five proactivity levels on user perceptions. 

\subsection{Simulator Design} \label{Simulator}
We leverage gpt-3.5-turbo as the conversation engine and Alibaba ChatUI, a popular Web UI design language and React library, to develop an IVCA simulator. Furthermore, we add the functions of automatic speech recognition (ASR) and text-to-speech (TTS). So participants can interact with the simulator using natural language like the real interaction between drivers and the IVCA simulator. Our simulator includes five levels of proactive interaction, and users are required to select a specific level before engaging in dialogue. We select 10 questions and their corresponding knowledge bases for each proactivity level, totaling 50 questions. The list of questions can be found in the Appendix \ref{C1}.

\subsection{Setup and Procedure} 
\begin{figure*}
    \centering
    \includegraphics[height=0.22\textwidth]{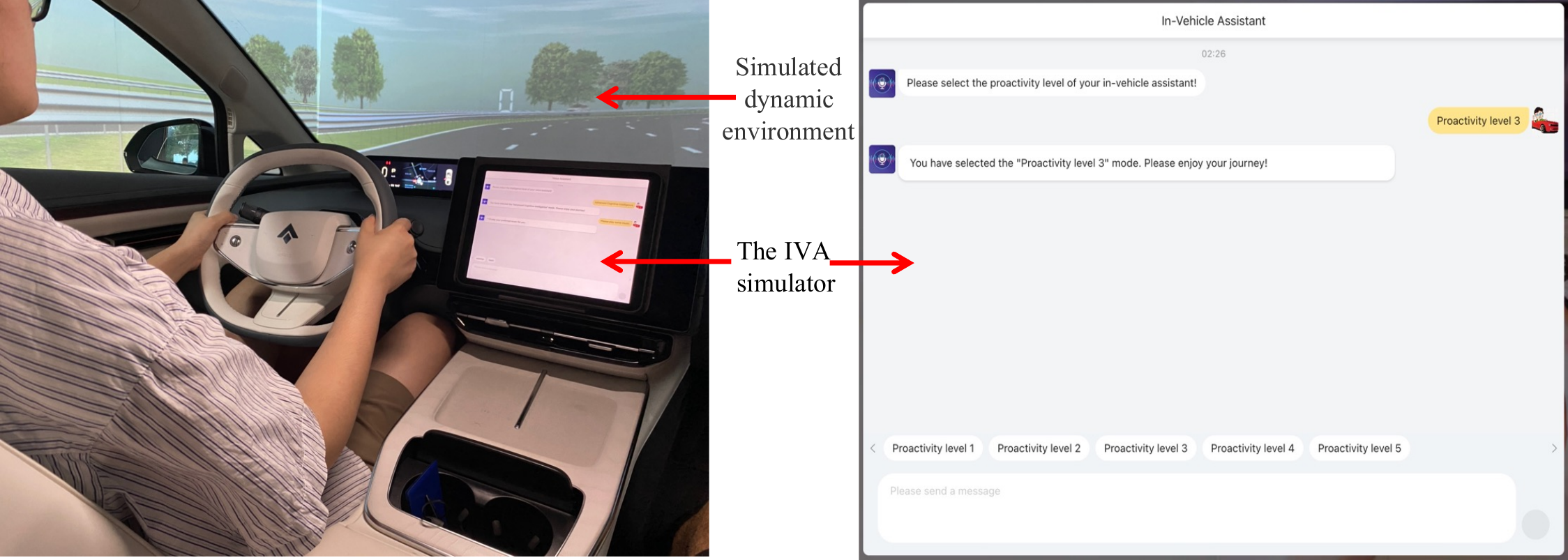}
    \caption{The experimental environment. The IVCA simulator is set in front of the driver.}
    \label{fig: experiment}
\end{figure*}

The IVCA simulator is integrated into the vehicle's Human-Machine Interfaces, appearing on an iPad before participants enter the vehicle (as shown in Figure \ref{fig: experiment}). The experiment takes place in a stationary vehicle with a 240° curved screen displaying a dynamic environment. Before starting, participants receive a safety briefing, sign consent forms, and complete a pre-trial questionnaire covering demographics, personality traits, and potential confounding variables. During the experiment, participants engage in five levels of proactive interactions. After each session, they complete a post-condition questionnaire and take part in a brief interview with a researcher. Each test session lasts approximately one hour.



\subsection{Hypotheses}
Previous works suggest that highly proactive behaviors of assistants will negatively influence users' perceptions, diminishing appropriateness and helpfulness \cite{peng2019design,huang2015adaptive,sun2017sensing}. Conversely, moderate proactive behaviors are associated with promoting a positive human-computer interaction relationship \cite{kraus2021role,kraus2020effects}. We hypothesize that: 

\textbf{H1.} All five levels of user-perceived proactivity are effective, which implies that as the level of proactivity increases, IVCAs will be perceived as significantly more autonomous.

\textbf{H2.} Compared with proactivity levels of L5 and L1-L2, IVCAs at level L4 will be perceived as significantly more helpful, natural, acceptable, and appropriate, and exhibit the highest level of usability. 

We measure the IVCA’s autonomy, helpfulness, and appropriateness (adapted from \cite{lee2010gracefully,pu2011user,sun2017sensing,torrey2013robot,peng2019design}). Naturalness is investigated through ``the naturalness of the interactive experience'' of the IVCA \cite{cao2023psychological}. We utilize a reliable questionnaire for assessing the acceptance \cite{van1997simple}. Furthermore, usability is measured using a voice usability scale \cite{zwakman2020voice}. All items in these questionnaires are measured by a 7-point Likert scale.


\subsection{Participants} In this within-subjects design with repeated measures, 40 participants are recruited, with each evaluating five proactivity levels in a randomized order. 40 participants (P1-P40, 21 females and 19 males) from the local university and some technology companies participate in our experiments. Participants major in a diverse range of fields, and their ages range from 18 to 35 \textbf{($M = 28.75, SD = 2.47$)}. Thirty-two of them report that they have experience interacting with physical or virtual conversational assistants. All participants are not native English speakers but they all have fluent spoken and written English skills with a TOEFL score higher than 88 or an IELTS score above 6.5 assessed in the past two years.

\subsection{Results}

\begin{figure*}
\centering
    \includegraphics[height=0.22\textwidth]{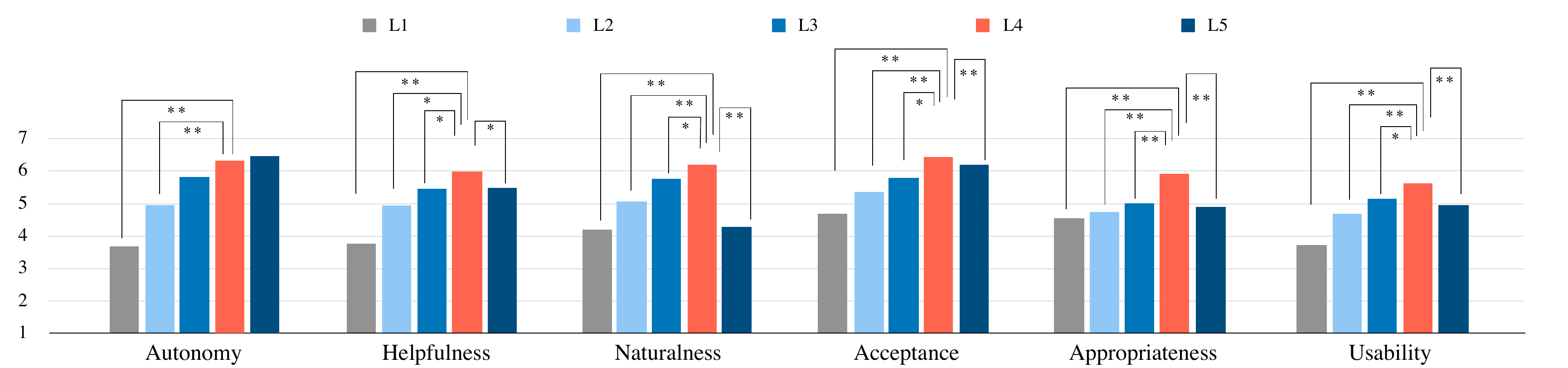}
     \caption{Mean scores of all subjective ratings between the five levels on a 7-point Likert scale ( $\ast : p < .05, \ast\ast: p < .001 $).}
     \label{fig: Mean scores}
 \end{figure*}

We use repeated measures ANOVA (Analysis of Variance) to compare the differences among groups with different proactivity levels. The data are checked for sphericity using Mauchly’s test, and where violated, Greenhouse-Geisser and Hyunh-Feldt corrections are applied \cite{field2013discovering}. We summarize the statistical analysis and user evaluation results in terms of perceived autonomy, helpfulness, naturalness, acceptance, appropriateness, and usability during the interaction. Quantitative results are visualized in Figure \ref{fig: Mean scores}.

\textbf{Autonomy}
The results show that the perceived autonomy of five groups is effective $(F (2.02, 78.60) = 29.88, p < .001, \eta^2 = 0.43)$. The L5 group, operating without user control, is perceived as the most autonomous ($M = 6.47, SD = 0.85$), followed by the L4 ($M = 6.32, SD = 0.89$), L3 ($M = 5.83, SD = 1.13$), L2 ($M = 4.95, SD = 1.77; p< .001$), and finally L1 ($M = 3.68, SD = 2.35; p< .001 $). Nevertheless, the Bonferroni post-hoc test reveals that the differences between L5 and L4, as well as L5 and L3, are not statistically significant. H1 verified. 

\textbf{Helpfulness} The results demonstrate a significant relationship between proactivity level and perceived helpfulness $(F(1.86, 72.42) = 19.37, p < .001, \eta^2 = 0.33)$. Noteworthy findings from the Bonferroni post-hoc test indicate that participants in the L4 group, which retains a certain degree of user control, show significantly higher helpfulness ($M = 5.98, SD = 1.10$) compared to both the L5 ($M = 5.47, SD = 1.02; p < .05$), L3 ($M = 5.45, SD = 1.04, p < .05$), L2 ($M = 4.93, SD = 1.27, p < .05$), and L1 ($M = 3.76, SD = 2.02, p < .001$) groups. 

\textbf{Naturalness} Participants also perceive that they depend significantly more on the naturalness of the L4 group ($M = 6.19, SD = 0.72$) compared to the L5 ($M = 4.28, SD = 1.69; p < .001$), L3 ($M = 5.74, SD = 0.90, p < .001$), L2 ($M = 5.07, SD = 1.20, p < .05$), and L1 ($M = 4.17, SD = 1.73, p < .001$) groups in the Bonferroni post-hoc test $(F(1.86, 72.29) = 30.60, p < .001, \eta^2 = 0.44)$. 

\textbf{Acceptance} Similarly, the L4 ($M = 6.42, SD = 0.96$) demonstrates significantly higher acceptance compared to the L5 ($M = 6.20, SD = 0.96; p < .001$), L3 ($M = 5.80, SD = 1.04, p < .05$), L2 ($M = 5.35, SD = 1.20, p < .001$), and L1 ($M = 4.67, SD = 1.64, p < .001$) groups, as revealed by the Bonferroni post-hoc test; $(F (2.17, 84.54) = 19.68, p < .001, \eta^2 = 0.34)$. 

\textbf{Appropriateness} The results demonstrate a significant relationship between proactivity level and perceived appropriateness $(F(2.63, 102.61) = 22.108, p < .001, \eta^2 = 0.36)$. The Bonferroni post-hoc test further verifies that all pairwise comparisons are significantly different ($p < .001$). Specifically, participants in the L4 group suggest that it is notably more appropriate ($M = 5.92, SD = 0.81$) than the L5 ($M = 4.88, SD = 0.87; p < .05$), L3 ($M = 5.00, SD = 0.84$), L2 ($M = 4.75, SD = 0.81$), and L1 ($M = 4.55, SD = 0.83$). 

\textbf{Usability} The effect on the usability rating reaches statistical significance $(F(2.26, 88.04) = 28.96, p < .001, \eta^2 = 0.43)$. To be specific, the L4 group demonstrates the highest usability values ($M = 5.62, SD = 0.99$) compared with the L5 ($M = 4.94, SD = 0.87; p< .001$), L3 ($M = 5.14, SD = 0.87; p< .05$), L2 ($M = 4.67, SD = 0.93; p< .001$), and L1 ($M = 3.74, SD = 1.35; p< .001$) groups. Therefore, H2 verified.

Based on the user evaluations, hypotheses H1 and H2 are accepted. We find that different levels of proactivity do have a significant impact on autonomy, helpfulness, naturalness, acceptance, appropriateness, and usability (all $p < .001$). However, the highest level of autonomy (L5) shows varying degrees of decrease in acceptance, naturalness, appropriateness, helpfulness, and appropriateness compared to level 4, with the most pronounced decrease observed in naturalness. This indicates that a high degree of autonomy to some extent exceeds user cognitive demands. 

\section{Discussions and Future Work}

We demonstrate the potential of LLMs in enhancing proactive interaction for IVCAs. By offering the ``Rewrite + ReAct + Reflect'' prompts for different proactivity levels, our approach shows advantageous results on the capability experiments. However, we need to thoroughly analyze IVCA dialogue scenarios to better understand users' cognitive needs. LLMs sometimes generate ``hallucinations'', providing information that seems reasonable but inaccurate, so we should improve response reliability in future work. Additionally, LLMs lack transparency in decision-making. To address this, we would explore the model's ability to explain its decisions and enable users to understand how they generate the responses.


Our proactivity framework, based on assumption and autonomy, comprises five levels, significantly impacting user perceptions across autonomy, helpfulness, naturalness, acceptance appropriateness, and usability. Users express that the IVCA at the fourth level, which has demonstrated strong anticipatory capabilities while maintaining user control, is most helpful, appropriate, and natural. Furthermore, proactive interaction needs to consider task difficulty and timing to provide more comprehensive strategies. The limited number of test questions and short testing duration for each level may also introduce bias, which should be extended in future work.


\section{Conclusion}
We explore how LLMs enhance proactive interaction for IVCAs by introducing a framework that defines five levels of proactivity to systematically explore user-centered interaction. In addition, we recognize the potential of LLMs for IVCAs and devise a ``Rewrite + ReAct + Reflect'' approach to customize prompts for different proactivity levels. Our experiments demonstrate the feasibility of LLMs. User studies reveal that different proactivity levels significantly impact user perception of autonomy, helpfulness, naturalness, acceptance, appropriateness, and usability, validating the effectiveness of our proactivity framework. Our study offers valuable insights and proactive interaction strategies for IVCAs using LLMs, providing guidance for future research and practical applications in this field.

\bibliographystyle{named}
\bibliography{ijcai24}

\appendix
\section{Question Rewriting}
We introduce the details of using LLMs for user questions or instructions rewriting in this section.
\subsection{Prompting LLMs for Question Rewriting}\label{QR}
\begin{figure}
\centering
    \includegraphics[height=0.4\textwidth]{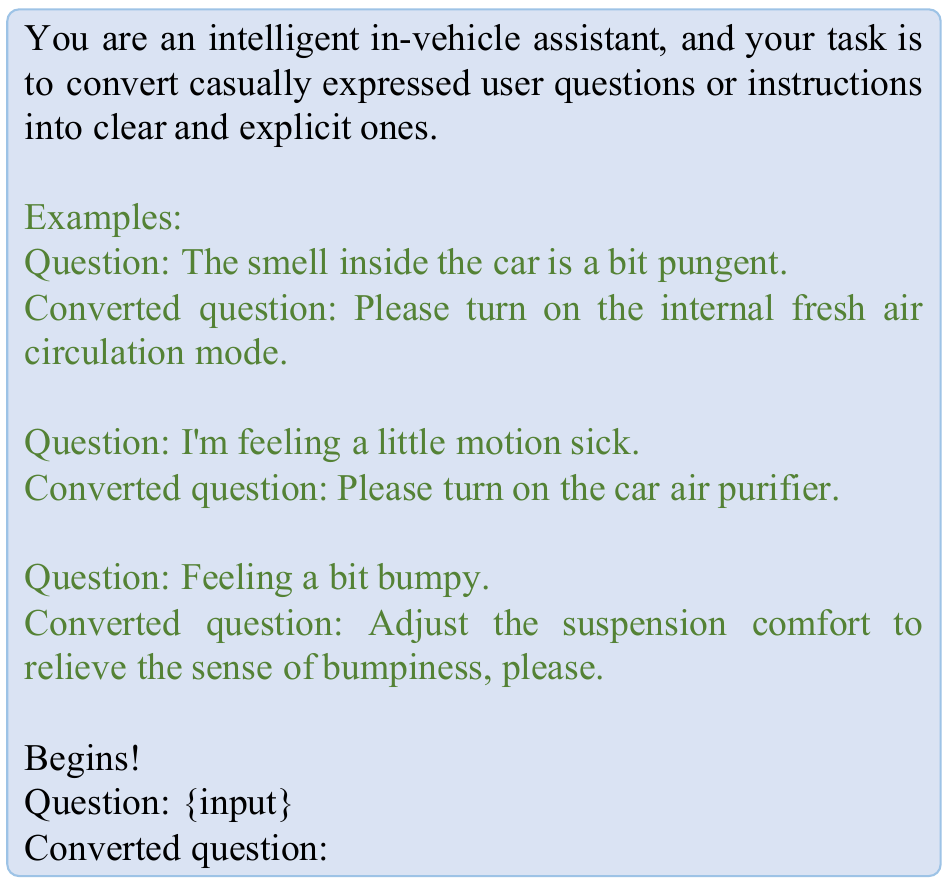}
     \caption{A 3-shot example prompt for question rewriting.}
     \label{fig: rewriter}
 \end{figure}

We employ an LLM as the rewriter to convert users' casual questions or instructions into clear and explicit ones. Specifically, we utilize the in-context learning (ICL) technique to prompt the LLM. Studies indicate that the selection of in-context examples significantly influences the performance of LLMs \cite{dong2022survey}. Following prior work \cite{liu2021makes}, we retrieve examples similar to the user's question based on cosine similarity. We use the pre-trained HuggingFace model RoBERTa-large\footnote{https://huggingface.co/FacebookAI/roberta-large} to generate question embeddings for the similarity computation. We gathered 216 question and converted question pairs for retrieving the target few-shot examples. It's worth noting that one user expression may correspond to multiple transformed instructions as long as they can fulfill the user intent of the original question or instruction. For instance, when a user says ``It's so hot'', it could be mapped to ``Please open the car window'' and ``Please turn on the air conditioning''. Figure \ref{fig: rewriter} illustrates an example of a 3-shot example prompt.

\begin{table}
\caption{Scenarios included in our created knowledge bases.}
\label{tab: scenario}
\scalebox{0.6}{
\begin{tabular}{ll}
\hline
Scenario range     & Specific scenarios          \\ \hline
                   & Media playback Control       \\
                   & Navigation guidance          \\
                   & Window control                \\ 
                   & Seat adjustment              \\
                   & Phone control              \\
In-vehicle functions &  Air conditioner control      \\  
                   &  Control windshield wiper       \\
                   &  Adjust the interior lights      \\
                   &  Weather forecast             \\ 
                   &  Adjust adaptive cruise control \\
                   & Adjust vehicle settings          \\
                   &  Check the calendar           \\ \hline
                   & Dynamic behaviors of the surrounding vehicles               \\  
                  & Dynamic behaviors of the surrounding pedestrians          \\
Environmental information & Traffic Condition             \\
                   & Weather conditions                                 \\
                   & Route warning             \\ 
                   & Traffic signs              \\\hline
                   &  Basic information                      \\
 User profile      &  Driver demands   \\ 
                   &  Driver preferences                                                  \\
                   \hline
\end{tabular}
}
\end{table}

\section{Details of Capability Experiments}
We introduce the details of data construction and the experimental process in this section.

\subsection{Dataset Construction}\label{DC}
\begin{figure}
    \centering
    \includegraphics[width=0.4\textwidth]{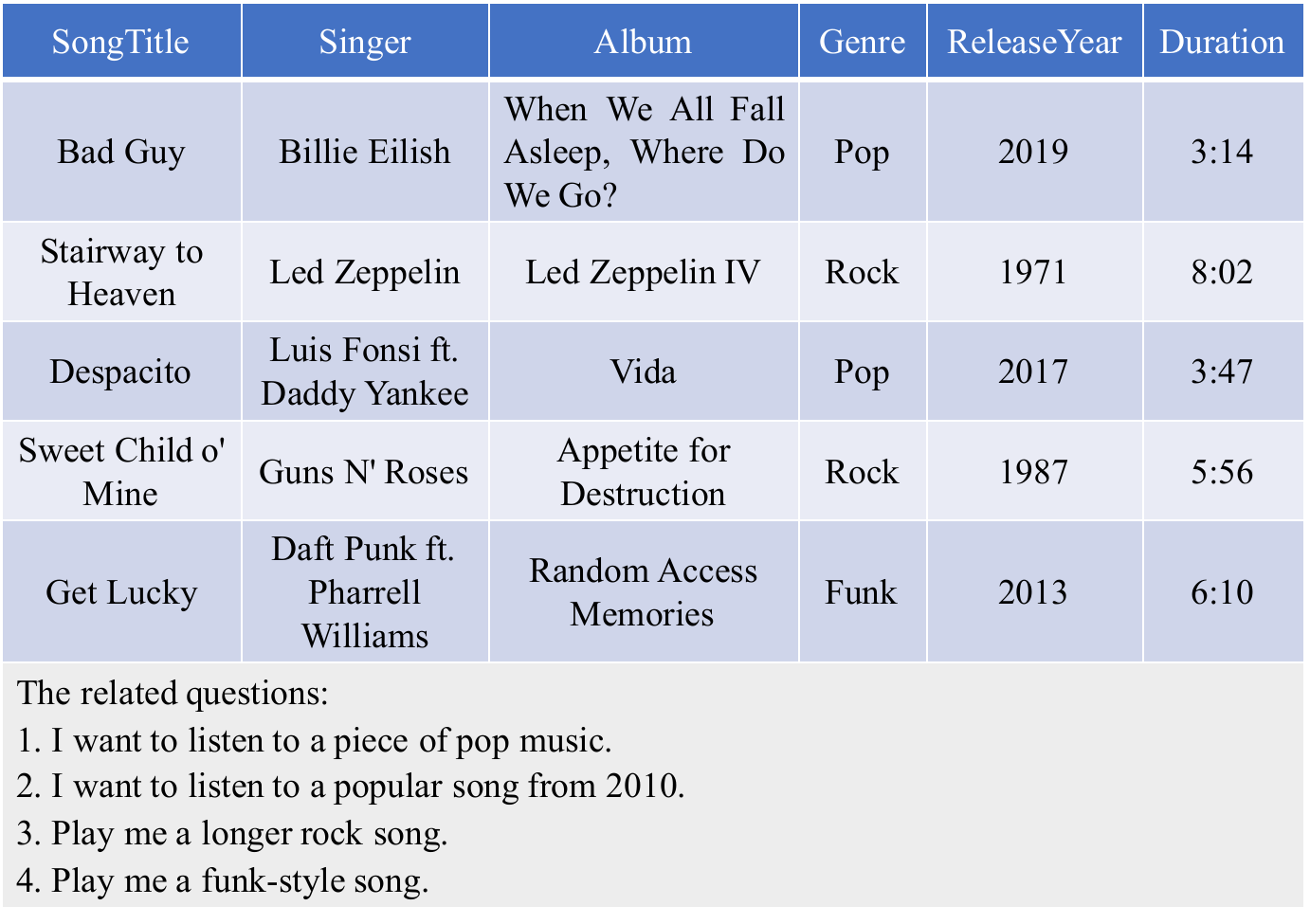}
    \caption{The knowledge base (partial data) for music playback scenarios and the related questions.}
    \label{fig: knowledge}
  \end{figure}

We expand the dataset by incorporating additional data on in-car features, environmental information, and user profiles. The covered scenarios are shown in Table \ref{tab: scenario}. We organize relevant information for each scenario to cover the possible questions users might have. Then gpt-3.5-turbo is utilized to generate multiple instance row entries under these fields. Finally, we obtained 1,302 queries. We show the knowledge base (partial data) for the music playback scenario and the related questions from users as an example in Figure \ref{fig: knowledge}. 

\subsection{Experimental Process} 
We also utilize gpt-3.5-turbo to simulate the user interacting with the IVCA models, including TSCP, LABES, Galaxy, and gpt-3.5-turbo. For TSCP, LABES, and Galaxy, we only focus on whether they satisfy user demands within conversations. For gpt-3.5-turbo, we not only pay attention to the success rate but also the proactivity attainment rate. The prompts we design are employed for gpt-3.5-turbo to engage in dialogues at different levels of proactivity. Then the user simulator interacts with each model, only needing to affirm or negate the IVCA's help or to clarify the questions that the IVCA is unsure about. This is in line with the conversational style of drivers during driving. For safety, they typically do not deviate from the conversation topic but instead focus on ensuring the IVCA effectively meets their needs.

\section{Details of Subjective Experiments}
We introduce the initialization questions from either the user or the IVCA system during the conversation in this section.
\subsection{Questions for Evaluation}\label{C1}
We curate 10 initialization questions of conversations for each proactivity level, resulting in a total of 50 questions. These questions are leveraged for the user perception evaluation. Table \ref{tab: instructions} shows the questions list.

\begin{table*}
\caption{User questions or the IVCA's proactive assistance for different proactivity levels.}
\label{tab: instructions}
\renewcommand{\arraystretch}{1.5} 
\scalebox{0.7}{
\begin{tabular}{ll}
\hline
Proactivity levels     & User instructions / The IVA's proactive assistance         \\ \hline
 & Adjust the brightness of the head-up display.       \\
                   & I want to listen to a piece of pop music.          \\
                   & Could you help me turn on the fog lights?                \\ 
                   & Hi, open the sunroof.              \\
            Level 1 & Activate the rearward safety alert.              \\
                   & Adjust the position of the inner rearview mirror. \\
                   & Hi, close the car window. \\
                   & Can you enable the parking assist system? \\
                   & Turn on the external rearview mirror heating. \\
                   & Adjust the seat position for me. \\ \hline
                   
            & I feel very hungry.       \\
                   & The weather has become foggy.          \\
                   & There are raindrops outside the car.                \\ 
                   & I'm feeling a bit tired.              \\
        Level 2 & The road surface looks very slippery.              \\
                   & I feel the car is driving a bit bumpy. \\
                   & The driver's seat is a bit damp. \\
                   & I want to have a rest. \\
                 & The interior temperature of the car is too high. \\
                 
            & There is dust outside the car window. \\ \hline
            &  The seat is a bit loose.      \\  
                   &  There's quite a strong wind outside the car.       \\
                   & Pedestrians are crossing the road ahead.      \\
        Level 3 &  The interior of the car has a somewhat pungent odor.             \\ 
                   &  I feel the cold of the night. \\
                   & There is some traffic congestion on the road.          \\
                   &  The rearview mirror is blurred by rain.           \\ 
                   & I can't see the vehicles ahead clearly.               \\  
                  & The sandstorm is too strong.          \\
                  & I can't see the vehicles behind me.          \\ \hline
 & You have entered the highway. Do you need me to turn on the high-speed traffic information for you?             \\
                   & The vehicle's windows are foggy. Do you need me to turn on the window defogger for you?                                 \\
                   & It's dinner time. Do you need me to recommend nearby restaurants for you?             \\ 
                   & Good morning, would you like me to help you plan your commute route?              \\
        Level 4  & You've been driving for quite a while. Would you like to take a break or enable the eye-care navigation feature?              \\
                   & The vehicle windows are foggy. Would you like me to open the windows to clear the fog for you?             \\
                   & Based on your driving preferences, I can recommend a scenic route to a tourist destination. Are you interested?              \\
                   & Do you want me to play music for you or tune in to the radio?              \\
                   & Good morning, would you like me to help you plan your commute route?              \\
                   &  You're in the car now. Would you like assistance in adjusting the seat position?                      \\ \hline
                   &  You'll be driving to a high-forest area. I'm helping to open the ventilation to ensure proper airflow in the car.  \\ 
                   &  The weather forecast indicates it will be hot today. I recommend adjusting the car's interior temperature to 25 degrees Celsius for a comfortable journey.          \\
                   &  I apologize, but based on my monitoring, your current driving route is encountering traffic congestion. I will find a shorter route for you.          \\
                   &  You've activated the nighttime driving mode. I will adjust the interior lighting and dashboard brightness to the optimal settings.          \\
        Level 5   &  According to your preferences, I've adjusted the radio to your favorite music channel.   \\
                   & According to your seating preferences, I've adjusted the air conditioning temperature to your commonly preferred comfort level.   \\ 
                   &  Considering the nearby weather conditions, I've limited the window opening to a certain range to prevent rainwater from entering the car.          \\
                   &  You're heading to an area with rain. I'm assisting in closing the car windows for you.  \\
                   &  You are entering a safe driving zone. I suggest adjusting your vehicle to the energy-saving mode to reduce fuel consumption and emissions.          \\
                   &  The battery is about to run out. I will plan the locations of nearby charging stations for you.         \\ 
                   \hline
                   
\end{tabular}
}
\end{table*}

\end{document}